\documentclass[sigconf,nonacm]{acmart}
\AtBeginDocument{%
  }
\usepackage{orcidlink}
\usepackage{url}
\usepackage{multirow}
\usepackage{array}
\usepackage{marvosym}
\usepackage{subcaption}
\usepackage[export]{adjustbox}

\setcopyright{none}
\begin{document}

\title{TenderKG}
\subtitle{A Knowledge Graph Dataset for Recommendation in Public Procurement\\ \ }


\author{Yacine Mokhtari \orcidlink{0009-0004-9080-9256}}
\affiliation{
    \institution{IMT Atlantique, Lab-STICC, UMR 6285, CNRS}
    \city{Brest}
    \country{France}
}
\affiliation{
    \institution{Intescia Group}
    \city{Issy-les-Moulineaux}
    \country{France}
}
\email{yacine.mokhtari@imt-atlantique.fr}

\author{Véra Pukhkoy}
\affiliation{
    \institution{Intescia Group}
    \city{Issy-les-Moulineaux}
    \country{France}
}
\email{vpukhkoy@intescia.com}

\author{Grégory Smits \orcidlink{0000-0002-0436-9273}}
\affiliation{
    \institution{IMT Atlantique, Lab-STICC, UMR 6285, CNRS}
    \city{Brest}
    \country{France}
}
\email{gregory.smits@imt-atlantique.fr}

\renewcommand{\shortauthors}{Mokhtari et al.}
\definecolor{darkblue}{RGB}{0, 51, 102}

\begin{abstract}
    Public procurement represents a major economic activity, where public institutions allocate contracts to companies through competitive tendering processes. 
    Despite its importance, this domain remains underexplored by recommender systems, largely due to the lack of publicly available datasets capturing its complexity.
    In this paper, we introduce TenderKG, a large-scale knowledge graph dataset constructed from French public procurement data covering the period 2021–2023. 
    The dataset models the procurement ecosystem through heterogeneous entities, including companies, tenders, lots, and domain-specific taxonomies of work domains, connected via rich semantic and structural relations. 
    A key specificity of this setting is that only the awarded companies are visible, resulting in sparse explicit signals of awarded interactions.
    To overcome this limitation, TenderKG integrates extensive side information on the actors in the French tender market and the tenders, including textual descriptions, hierarchical classifications, and geographical features, enabling the study of knowledge-aware recommendation in a highly constrained and competitive environment. 
    We provide detailed statistics and analyses of the dataset, highlighting its structural properties, sparsity patterns, and domain-specific characteristics.
    We believe TenderKG opens new research directions in bidder recommendation, knowledge graph-based recommendation, competition-aware matching, and provides a valuable benchmark for evaluating methods in real-world, high-stakes decision-making scenarios.
\end{abstract}




\maketitle

\section{Introduction}
\begin{figure}[t]
    \centering
    \includegraphics[width=0.95\linewidth]{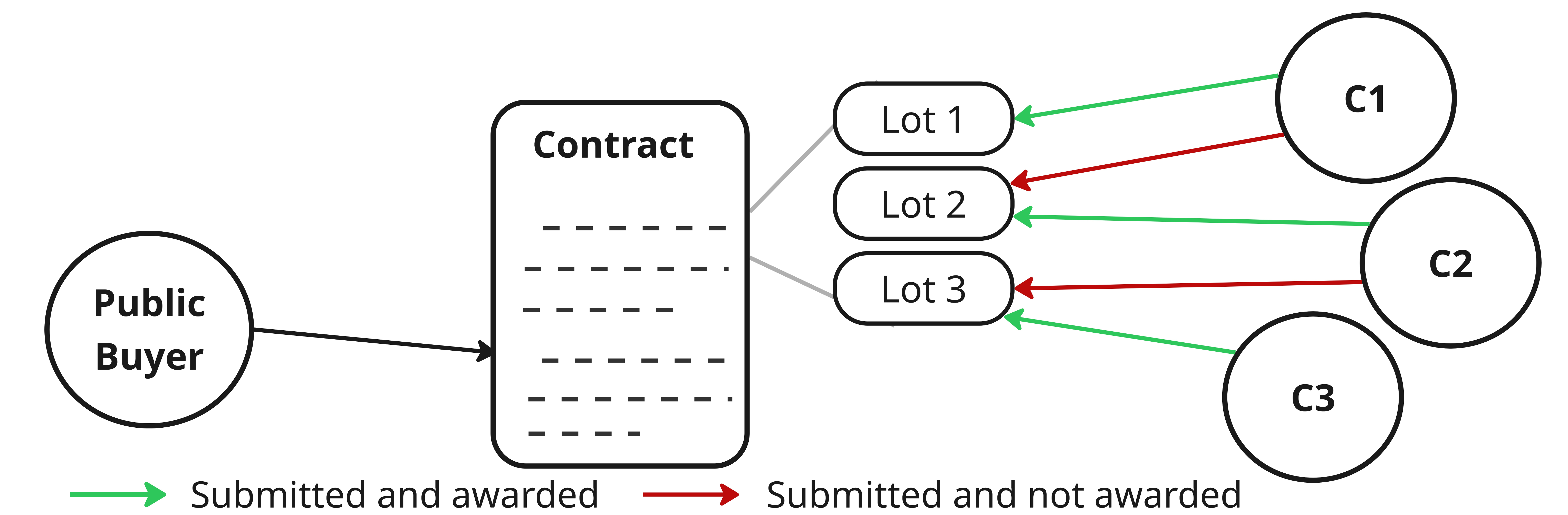}
    \caption{Illustration of the procurement process, showing interactions between public buyers, contracts, lots, and bidding companies (C).}
    \label{fig:introduction_public_procurement}
\end{figure}

Public procurement refers to the process by which public institutions acquire works, goods, and services from external companies~\cite{garcia2020bidders}. 
Representing one of the largest sources of economic activity in many countries\footnote{\href{https://single-market-economy.ec.europa.eu/single-market/public-procurement/digital-procurement/public-procurement-data-space-ppds\_en}{The Public Procurement Data Space (PPDS)}}, public procurement plays a crucial role for both companies seeking business opportunities~\cite{garcia2020bidders} and researchers interested in large-scale economic ecosystems.

Public procurement processes are typically formalized through announcements known as \textit{public tenders}, issued by public agencies (hereafter referred to as \textit{public buyers}). Each tender is generally divided into multiple \textit{lots}, corresponding to specific sub-parts of the contract. \textit{Bidding companies} can submit bids to one or several lots, and ultimately, a single company, referred to as the \textit{beneficiary}, is selected for each lot (see Fig.~\ref{fig:introduction_public_procurement}).

In recent years, there has been a growing interest in developing bidder recommender systems\cite{garcia2020bidders, nai2023public}, which aim to assist either (1) public buyers in identifying suitable companies for their tenders, or (2) companies in discovering relevant bidding opportunities. These systems operate in a particularly challenging setting.
First, interactions between companies and lots are highly sparse. When made publicly available, the number of submissions range under 5 companies per lot \cite{nai2023public}. 
Second, interactions alone are far from sufficient. The recommendation process involves rich contextual and domain-specific features: On the one hand, tender characteristics, such as amount, type, lot nature, public buyer, location, among others, and, on the other hand, company characteristics including activity sector and staffing range.
Third, the problem is inherently competitive, each lot can only be awarded to a single company, and asymmetric, both parties (public buyers and bidding companies) have distinct preferences and constraints.

These characteristics distinguish bidder recommendation from other well-studied recommendation scenarios such as e-commerce or multimedia. 
One can observe similarities with reciprocal recommendation, while also introducing unique constraints related to competition and exclusivity that can occur in job hunting scenarios. 
As a result, modeling this problem solely as a user-item bipartite graph is insufficient. Instead, capturing the full ecosystem requires leveraging structured domain knowledge to incorporate heterogeneous relationships and high-level semantic information.

Despite studies on public tenders~\cite{deschamps2025processing}, to the best of our knowledge, a well-structured dataset with rich side information tailored for Tender Recommendation does not publicly exist.

To address this gap, we present \textit{TenderKG}, a Knowledge Graph (KG) dataset derived from French public procurement data, collected and processed through an industrial pipeline developed by Intescia Group\footnote{\url{https://intescia-group.com}}. The dataset covers public tenders published between [2021–2023], including public buyers, tenders, lots, and awarded companies, along with domain-related side information.
The consolidated dataset is made publicly available at FigShare\footnote{\textcolor{darkblue}{\url{https://figshare.com/articles/dataset/TenderKG_A_Knowledge_Graph_Dataset_for_Recommendation_in_Public_Procurement/32149909}}}. Furthermore, the Intescia Group will provide updates to the dataset, enabling longitudinal and trend analyses.

Our contributions can be summarized as follows:
\begin{itemize}
    \item \textbf{Curated multi-sources dataset:} We provide a dataset processed from multiple heterogeneous public sources, involving extensive data cleaning, normalization, and entity resolution.
    \item \textbf{Domain-specific KG:} We release a structured representation of the procurement ecosystem, capturing key entities and relationships, including hierarchical taxonomies such as CPV (Common Procurement Vocabulary) and NAF (activity sectors).
    \item \textbf{Dataset characterization:} We present a comprehensive analysis of the dataset, including a concise overview of the procurement domain and key statistical properties (e.g., distributions, sparsity, and heterogeneity), highlighting challenges for recommendation tasks.
\end{itemize}
    
We believe that this dataset opens new research directions for recommendation systems in highly competitive and knowledge-rich environments. In particular, it enables the study of competition-aware recommendation, semantic matching, and graph-based learning approaches.
Since only awarded submissions accounts as user interactions, this enables reflecting on beyond accuracy metrics tailored for high-stakes decision-making scenarios.




\section{Dataset Description}

\subsection{Collection Process}
The TenderKG dataset comprises all awarded French public tenders over a three-year period (2021-2023), collected from multiple official sources: including national and European publication platforms (e.g., BOAMP\footnote{\href{https://www.boamp.fr/pages/entreprise-accueil/}{https://www.boamp.fr/pages/entreprise-accueil/}}, JOUE\footnote{\href{https://eur-lex.europa.eu/oj/direct-access.html?locale=en}{https://eur-lex.europa.eu/oj/direct-access.html?locale=en}}), as well as regional procurement portals.
As the dataset is constructed from real-world contract award notices, its quality is inherently tied to the reporting practices of public buyers, and may therefore exhibit variability and missing information.
The data has been prepared through an industrial collection and processing pipeline developed by Intescia Group, involving document retrieval, information extraction and entity resolution across heterogeneous sources. 

As this work focuses exclusively on publicly available procurement data in France, the dataset is released without anonymization. 
It covers a wide range of companies, from very small bidding companies (1-2 employees) to large organizations with more than 10 000 employees, spanning diverse industry sectors.

A key characteristic of this dataset is that only the beneficiaries (i.e., awarded companies) are publicly disclosed. 
Unsuccessful bidders are, on the other hand, not disclosed by public buyers.
As a result, observed interactions between public buyers and bidding companies are extremely sparse and limited to positive outcomes. 
This setting poses significant challenges for recommender systems relying solely on collaborative signals, as the underlying interaction graph is highly constrained.
To mitigate this limitation, we also provide rich tender-side and company-side information, when available, including tender titles and descriptions, awarded amounts, hierarchical tender classifications (official CPV taxonomy), company activity sectors (official NAF), and geographical information such as departments, neighboring relations, and regions.


\subsection{Structure and Content}
\begin{figure}[t]
    \centering
    \includegraphics[width=0.94\linewidth]{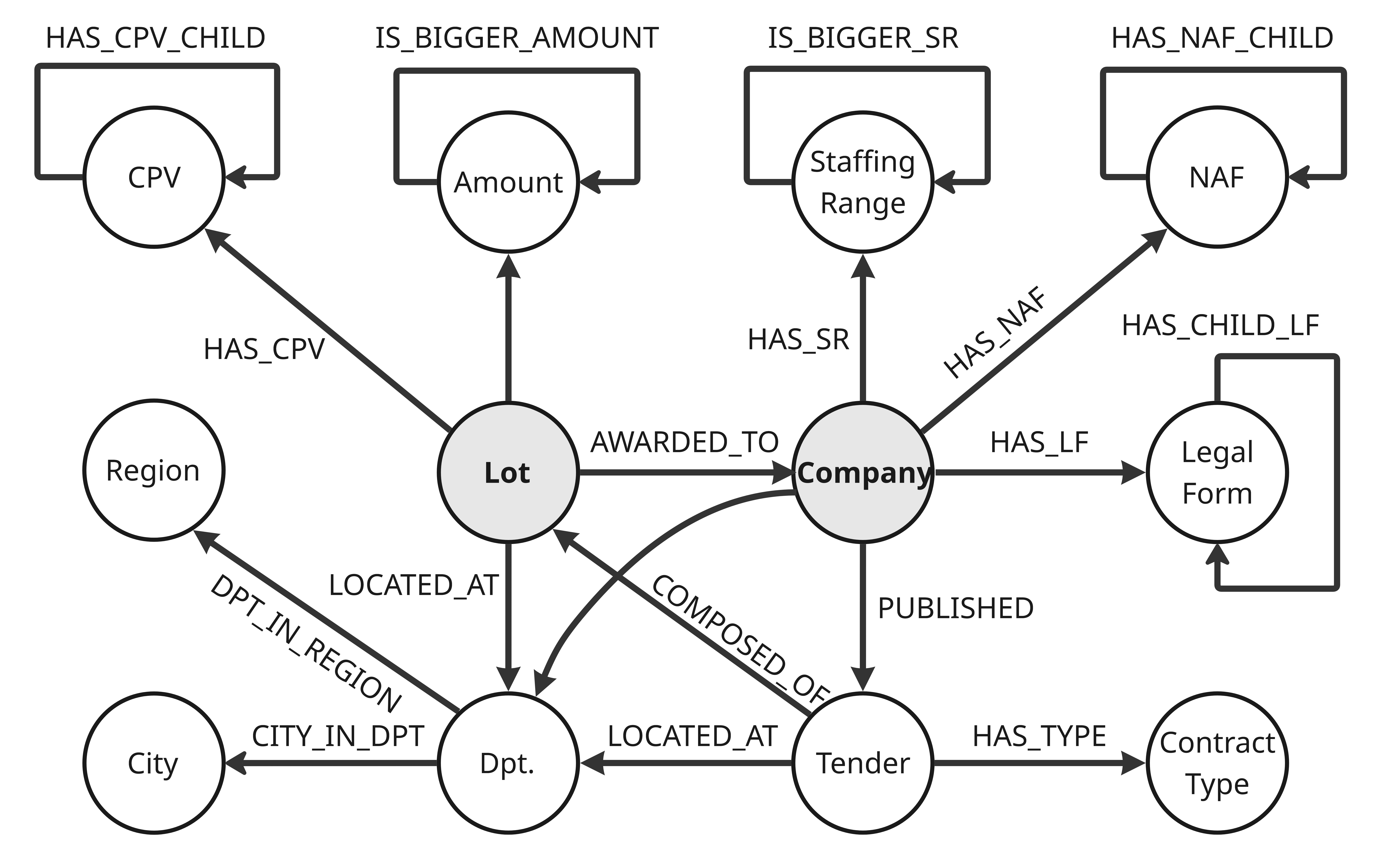}
    \caption{TenderKG Schema}
    \label{fig:ontology}
\end{figure}
The TenderKG dataset is structured as a KG, where information is represented as a collection of factual triplets of the form {\it (head, relation, tail)}. The overall schema of the KG is illustrated in Fig.~\ref{fig:ontology}, comprising 12 entity types and 21 relation types. The summary of TenderKG statistics are reported in Table~\ref{tab:node_properties}.

\begin{table}[t]
\centering
\caption{Summary of dataset statistics. \textbf{B} and \textbf{PB} denote Beneficiaries and Public Buyers, respectively. \textbf{Dual-role} refers to companies that act as both beneficiaries and public buyers in TenderKG. CPV codes are reported at the 5th level.}
\label{tab:node_properties}

\small
\setlength{\tabcolsep}{5pt} 

\begin{tabular}{l c l c | l c}
\toprule
\textbf{Object} & \textbf{Count}
& \textbf{Object} & \textbf{Count}
& \textbf{Objects} & \textbf{Sparsity} \\
& &  & &  & \textbf{ratio} \\
\midrule
Tenders & 6,126
& Companies & 15,340
& B--CPV & 99.62\% \\ 

Lots & 29,220
& \hspace{0.8em}B & 13,033
& NAF--CPV & 99.45\%\\

CPV & 2,468
& \hspace{0.8em}PB & 2,359
& \\

NAF & 732
& \hspace{0.8em}Dual-role & 52
&  &  \\

\bottomrule
\end{tabular}
\end{table}


Below, we describe the main entity types and their associated semantics:\\

    \textbf{Company:} This represents both beneficiaries and public buyers, as they share common attributes. A company is identified by a unique identifier (SIREN) and is associated with metadata such as name, registration date, and, when publicly available, financial indicators. \\
    \textbf{Tender:} Corresponds to a procurement announcement issued by a public buyer. It constitutes the main document in the dataset and includes a unique identifier, a short title, and a textual description, typically consisting of one to three sentences summarizing the procurement objective. \\
    \textbf{Lot:} Each tender is decomposed into one or several lots, representing finer-grained procurement units. Lots are associated with textual information (title and description) and correspond to the actual items awarded to companies. \\
    \textbf{CPV (Common Procurement Vocabulary):} CPV codes follow a tree structure system defined by the European Union\footnote{\href{https://single-market-economy.ec.europa.eu/single-market/public-procurement/digital-procurement/common-procurement-vocabulary\_en}{CPV: https://single-market-economy.ec.europa.eu/single-market/public-procurement/digital-procurement/common-procurement-vocabulary\_en}}. They are 9-digit hierarchical identifiers, where the first digits define broad categories and subsequent digits refine the classification. In practice, codes may be assigned at varying levels of granularity, and multiple CPVs can be associated with a single lot. We provide the full hierarchy and labels, enabling structural and semantic reasoning. \\
    \textbf{NAF (French Activity Classification):} Company activity sectors are described using the NAF classification defined by INSEE\footnote{\href{https://www.insee.fr/en/information/8621640}{NAF: https://www.insee.fr/en/information/8621640}}. This hierarchical taxonomy provides sector labels at multiple levels of granularity (levels 1-5), enabling the modeling of similarities between companies based on their activity domains. In contrast to CPV, each company (public buyer or beneficiary) is mapped to a unique leaf node. \\
    \textbf{Contract Type:} Encodes the four types of procurement, i.e., services, goods, works, and furniture, providing a coarse-grained categorization of tenders. Unlike lots that can be assigned to multiple CPVs, a tender belongs to only one contract type. It is also worth noting that a Contract Type does not have its own distinct set of CPVs; the same CPV can appear in lots from different tenders with distinct types.\\
    \textbf{Legal Form:} Represents the legal structure of companies, e.g., commercial company, public entity, which can influence eligibility and participation in tenders. \\
    \textbf{Staffing Range:} Companies are associated with discretized employee count ranges based on INSEE standards\footnote{\href{https://www.insee.fr/fr/information/1896455}{Last table on https://www.insee.fr/fr/information/1896455}}. These ranges are organized \textit{sequentially} to capture relative company sizes with the \texttt{IS\_BIGGER\_SR} relation. \\
    \textbf{Department, Region and City:} The dataset incorporates three levels of geographical granularity. Departments represent the primary administrative division in France and are associated with both codes and labels. Regions group multiple departments, while cities provide finer-grained localization. Additionally, departments are linked through neighboring relations, capturing geographical proximity, which are relevant for procurement decisions.\\
    \textbf{Amount:} While the exact awarded amount (in \EUR{}) is stored as a data property of lots, we additionally introduce a discretized representation into monetary ranges. These ranges, defined in collaboration with domain experts, include categories such as \textit{Very Low} (<10K), \textit{Low} (10K--50K), \textit{Medium Low} (50K--100K), \textit{Medium} (100K--300K), \textit{Medium High} (300K--500K), and \textit{High} (500K--1M). Similar to staffing ranges, these categories are organized hierarchically through \texttt{IS\_BIGGER\_AMOUNT} relations, enabling ordinal reasoning.\\

\begin{figure}[t]
    \centering
    \includegraphics[width=0.99\linewidth]{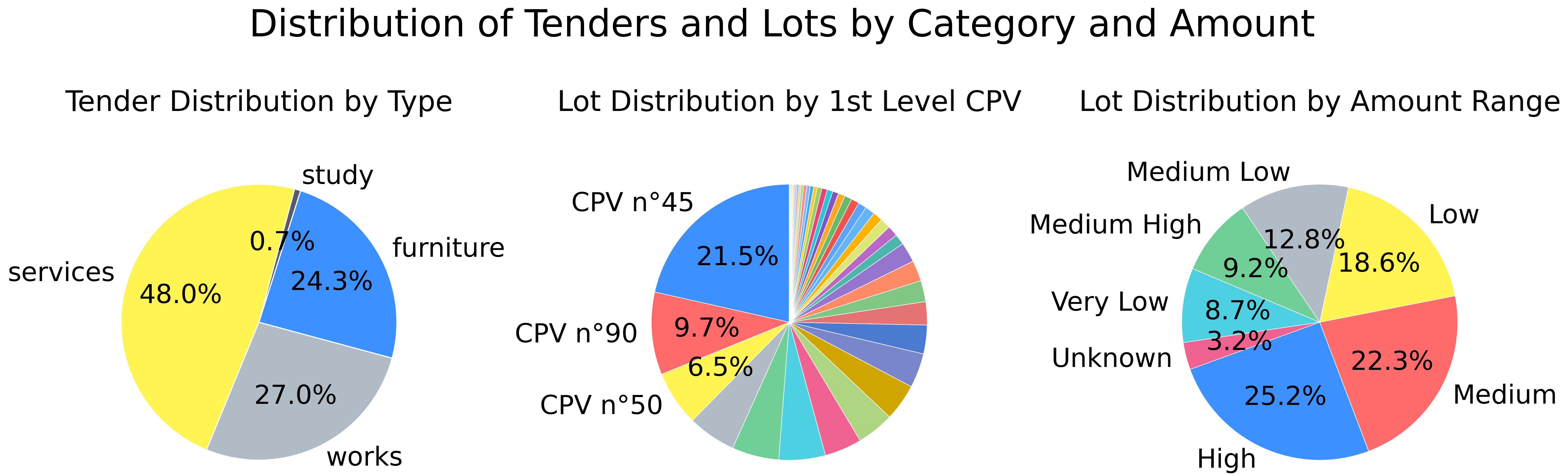}
    \caption{Distribution of Tenders and Lots by Type, CPV Category, and Amount Range in TenderKG. This diversity showcases the richness and heterogeneity of TenderKG.}
    \label{fig:contract_type_distribution}
\end{figure}

Fig.~\ref{fig:contract_type_distribution} presents the distribution of tenders by contract type (left), as well as the distribution of lots by 1st-level CPV category (center) and by amount range (right).
Most tenders correspond to \textit{Services}, followed by \textit{Works}, \textit{Supplies}, and \textit{Studies}. Regarding the lot distribution, as previously discussed, public buyers may assign CPV codes at varying levels of granularity. 
To ensure consistency and structured analysis, all CPVs are therefore aggregated to their 1st-level categories in Fig.~\ref{fig:contract_type_distribution}, resulting in 45 distinct groups. The resulting distribution highlights the diversity of TenderKG in terms of activity sectors. In particular, CPV categories 45 (\textit{Construction work}), 90 (\textit{Sewage, refuse, cleaning, and environmental services}), and 50 (\textit{Repair and maintenance services}) are the most represented, accounting for $21.5\%$, $9.7\%$, and $6.5\%$ of all lots, respectively.
Finally, the distribution across amount ranges shows that most lots fall within the \textit{High} ($25.2\%$) and \textit{Medium} ($22.3\%$) ranges, while only a small proportion ($3.2\%$) corresponds to lots for which no amount information is provided by public buyers.
Overall, these distributions highlight the heterogeneity of TenderKG in terms of contract types, activity sectors, and economic scale.

\begin{figure}[t]
    \centering

    \begin{subfigure}[t]{0.45\linewidth}
        \centering
        \includegraphics[width=\linewidth]{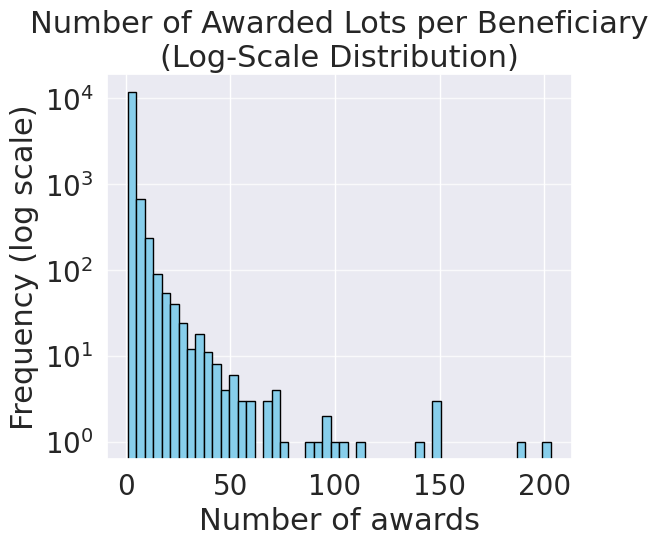}
        \caption{Distribution of the number of awarded lots per beneficiary.}
        \label{fig:awards_per_beneficiary}
    \end{subfigure}
    \hfill
    \begin{subfigure}[t]{0.45\linewidth}
        \centering
        \includegraphics[width=\linewidth]{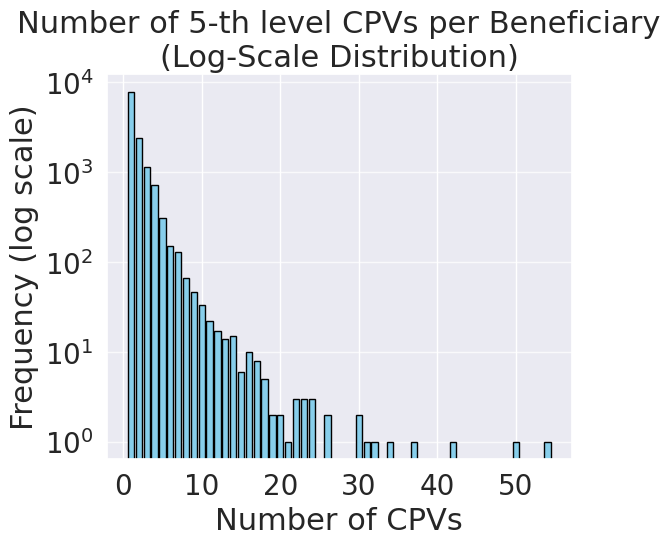}
        \caption{Distribution of the number of CPV categories per beneficiary.}
        \label{fig:cpv_per_beneficiary}
    \end{subfigure}

    \caption{Distribution of beneficiary activity in TenderKG. Both distributions exhibit a strong right-skew, indicating that most beneficiaries are associated with a small number of awards and categories, while a limited number of beneficiaries are highly active.}
    
    \label{fig:beneficiary_distributions}
\end{figure}

Fig.~\ref{fig:beneficiary_distributions} presents the distributions of beneficiary activity. Particularly, Fig.~\ref{fig:awards_per_beneficiary} shows the number of awarded lots per beneficiary. However, interactions at the lot level are binary (i.e., each award is observed only once), limiting the amount of collaborative signal.
To better characterize beneficiary activity, we instead consider their CPV coverage, i.e., the number of distinct 5th-level CPV categories associated with each beneficiary. This level is chosen as a trade-off between sparsity reduction (higher-level aggregation reduces the number of distinct branches) and semantic granularity. The resulting distribution (Fig.~\ref{fig:cpv_per_beneficiary}) follows a similar long-tailed pattern.

\begin{figure}
    \centering
    \includegraphics[width=0.77\linewidth]{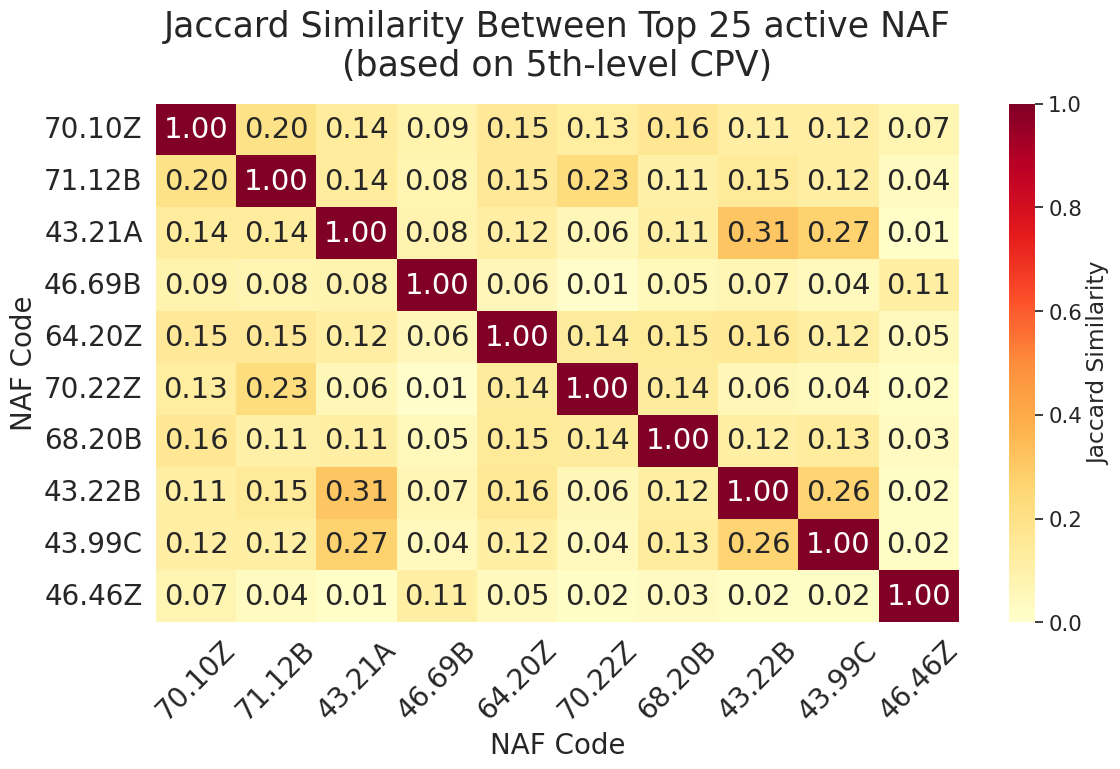}
    \caption{Jaccard similarity between NAF categories based on CPV assignments. The matrix highlights both intra- and cross-sector overlaps, showing that similarity is not strictly driven by the NAF hierarchy.}
    \label{fig:NAF_jaccard_similarity}
\end{figure}

While NAF codes provide a structural grouping of beneficiaries, Fig.~\ref{fig:NAF_jaccard_similarity} shows that proximity in the NAF hierarchy does not necessarily imply higher overlap in CPV assignments. In particular, some distant NAF categories exhibit comparable or higher similarity than structurally close ones, indicating that shared activity patterns are not strictly aligned with the taxonomy.
For instance, NAF categories 43.21A and 43.99C exhibit a relatively high similarity (0.27), while categories within the same branch such as 70.10Z and 70.22Z show lower overlap (0.13). Notably, 70.10Z is more similar to a category from another branch (68.20B), further indicating that shared activity patterns are not strictly aligned with the taxonomy.

\subsection{Potential Research Directions}
First experiments have been conducted on TenderKG using proven GNN-based recommenders~\citep{KGAT19,sun2020multi}. Due to the high sparsity and crucial role of side information, the obtained performances were significantly lower than those published in the papers. We therefore consider TenderKG as a challenging benchmark that motivates the development of models adapted to such conditions. The following potential research directions are opened up by TenderKG:
\begin{itemize}
    \item \textbf{KG + Textual representation learning:} Leveraging both hierarchical entities (e.g., CPV, NAF), and ordinal entities (such as Amount and Staffing Range) for entity representation.
    \item \textbf{Recommendation under sparsity and domain heterogeneity:} Addressing the lack of shared interactions and the diversity of activity sectors through side information (e.g., CPV, NAF), trading between grouping and specializing. Exploiting similarities across sectors and attributes to recommend opportunities for low-interaction beneficiaries.
    \item \textbf{Domain-specific recommendation:} Tacking into account domain-specific dimensions centered on buyer–supplier\footnote{A supplier is another term for a bidding company.} relationships~\cite{shamsollahi2021buyer}. When considering the competition between bidding companies, a poor preference match regarding a tender may be offset by lower competitiveness.
\end{itemize}

We argue that recommender systems in this context should balance between \textit{relationship continuity}, where entities have previously interacted, and \textit{relationship exploration}, where new partnerships must be discovered. This duality reflects fundamental dynamics of buyer–supplier interactions and poses additional challenges for recommendation.

\section{Conclusion}
The TenderKG dataset introduces a domain-rich dataset for public procurement, integrating tender-related information, and company side information within a unified KG. Its construction relies on extensive data collection, cleaning, and integration from multiple heterogeneous public sources by Intescia Group.

Through our analysis, we highlight key characteristics of the dataset, including strong sparsity, long-tailed distributions, and the presence of rich structural and semantic information. These properties make TenderKG a challenging and realistic benchmark for recommendation tasks in structured domains and high-stakes decision-making scenarios. In particular, the dataset opens research directions involving deep hierarchical reasoning, ordinal relations, and the integration of textual features. These research directions are the focus of ongoing work.

\bibliographystyle{ACM-Reference-Format}
\bibliography{sample-base}

\end{document}